%% file: main.tex
\documentclass[a4paper,conference]{IEEEtran}

\input{preamble}

\begin{document}
\title{Using Convolutional Codes for Key Extraction in SRAM Physical Unclonable Functions}

\author{\IEEEauthorblockN{Sven M\"uelich\IEEEauthorrefmark{1},
Sven Puchinger\IEEEauthorrefmark{1},
Martin Bossert\IEEEauthorrefmark{1}}
\IEEEauthorblockA{\IEEEauthorrefmark{1}
              Ulm University, 
              Institute of Communications Engineering, 
              89081 Ulm, Germany\\
              Email: \{sven.mueelich, sven.puchinger, martin.bossert\}@uni-ulm.de}
}

\maketitle

\begin{abstract}
Physical Unclonable Functions (PUFs) exploit variations in the manufacturing process to derive bit sequences from integrated circuits, which can be used as secure cryptographic keys. 
Instead of storing the keys in an insecure, non-volatile memory, they can be reproduced when needed. 
Since the reproduced sequences are not stable due to physical reasons, error correction must be applied. 
Recently, convolutional codes were shown to be suitable for key reproduction in PUFs based on SRAM. 
This work shows how to further decrease the reconstruction failure probability and PUF implementation size using codes with larger memory length and decoding concepts such as soft-information and list decoding.
\end{abstract}


%-----------------------------------------------------------------------
% Motivation
%-----------------------------------------------------------------------
\section{Motivation and Preliminary Work}

Generating cryptographic keys by using Pseudo Random Number Generators (PRNGs) and storing them in non-volatile memory has several drawbacks. 
First, PRNGs do not provide true randomness and hence violate the demand for unique and unpredictable keys. 
Second, storing keys makes them vulnerable to physical attacks and hence requires expensive protecting mechanisms which need additional chip area and power consumption. 
Physical Unclonable Functions (PUFs) are devices which extract unique cryptographic keys from intrinsic randomness like delay characteristics of digital circuits  or initialization behavior of memory cells. 
All devices (even with the same functionality) differ regarding this intrinsic randomness, which is unique for each device and cannot be controlled during the manufacturing process due to technical and physical limitations. A PUF uses this randomness to extract a random, unique and unpredictable sequence of bits which can be used as cryptographic key. 
Since the randomness is reproducible, the key can simply be regenerated when needed instead of storing it permanently.
Hence, PUFs implement secure key generation and secure key storage.

Since derived bit sequences suffer from environmental conditions like temperature or voltage supply, responses may vary slightly.
To circumvent this problem, \emph{error-correcting codes} have to be used.
Therefore, the initial PUF response is used to obtain \emph{helper data}, which is needed for reliable key reproduction.
Since enough uncertainty remains when knowing the helper data, a protected memory is not required.
Extraction of helper data and using them to reproduce a key can be done by different \emph{helper data schemes}, cf. \cite{linnartz2003new,dodis2004fuzzy,hiller2013breaking,muelich2016new}.

Previous work most often used concatenated schemes for error correction.
\cite{bosch2008efficient} suggested different ordinary concatenated schemes consisting of a repetition code as inner code and BCH, Reed--Muller and Golay Codes respectively as outer code.
\cite{maes2012pufky} applied an ordinary code concatenation consisting of a BCH and a repetition code.
Generalized concatenated codes were applied for the first time to PUFs in \cite{muelich2014error} using Reed--Muller codes, while a similar construction using BCH and Reed--Solomon codes was given in \cite{puchinger2015error}.

In \cite{hiller2013breaking}, convolutional codes were used the first time for error correction in PUFs.
Decoding was performed using a hard-decision Viterbi algorithm.
As channel model, a \emph{binary symmetric channel (BSC)} with variable crossover probabilities $\pi$ was assumed, where $\pi$ corresponds to the probability of a bit flip in the $i$-th SRAM cell.
In order to increase the key reproduction reliability, unreliable SRAM cells were ignored, i.e., for a given threshold $\pT$, only cells with $\pi < \pT$ were used, resulting in an implementation overhead.

In \cite{hiller2016} the decoding failure rate was further decreased using an estimator for the decoding output reliability (simplified ROVA) in combination with multiple readouts.

In this paper, we show that using full soft information at the Viterbi decoder's input, the decoding failure rate can be reduced significantly in the considered scenario.\footnote{Full soft information at the decoder's input was used for the first time in \cite{maes2009low}. However, instead of convolutional codes and Viterbi algorithm, a concatenation of short block codes was used. The use of convolutional codes was labeled inefficient without further study, due to apparently required very long code sequences. In contrast to that statement, \cite{hiller2013breaking,hiller2016} as well as our analysis show that sufficiently short length can be achieved.}
Furthermore, more unreliable bit positions can be included without losing error correction capability compared to \cite{hiller2013breaking,hiller2016}.
Thus, the decoding overhead vanishes.
We show, that the use of list decoding further improves the scheme in terms of failure probability and provides decoding reliability for free as an alternative to simplified ROVA. 
Also, the combination of list decoding and multiple readouts can improve the results of \cite{hiller2016}.%, by taking the intersection of the corresponding lists. 

Our simulations indicate that convolutional codes with larger memory length result in smaller failure probabilities, which is a well-known fact in coding theory. 
In order to make decoding of codes with a memory length up to 25 feasible, we apply sequential decoding.
To our knowledge, sequential decoding is applied to the PUF scenario for the first time.

\section{Physical Unclonable Functions (PUFs)}
\label{subsec:SRAM}
There is a variety of ways to realize PUFs. 
In this work we mainly focus on SRAM PUFs which were introduced in \cite{guajardo2007fpga}.
Dependent on the mismatch of two inverters of an SRAM cell (which is caused by manufacturing variations) and electrical noise, every SRAM cell converges to either 0 or 1 when powering-up the device.
Since the mismatch is static and dominates over the noise, most cells always power-up with either 0 or 1.
However, some of the cells do not have a preferred behavior and hence sometimes initialize with 0 and sometimes with 1.  
It is possible to produce reliability information of the individual cells by obtaining statistics during an enrollment phase.
Based on such statistics, \cite{maes2009soft} derived a cumulative density function (cdf) and a probability density function (pdf) which approximate the measured one-probability $P_{one}$ of SRAM cells:
\begin{align}
\cdf_{P_{one}}(x) = \Phi(\lambda_1 \cdot \Phi^{-1}(x)-\lambda_2)
\label{eqn:cdfpone}
\end{align}
\begin{align}
\pdf_{P_{one}}(x) = \frac{\lambda_1 \cdot \varphi(\lambda_2-\lambda_1 \cdot \Phi^{-1}(x))}{\varphi(\Phi^{-1}(x))}
\label{eqn:pdfpone}
\end{align}
In Equations~(\ref{eqn:cdfpone})--(\ref{eqn:pdfpone}), $\Phi$ denotes the standard normal cumulative distribution function.
For these functions we use $\lambda_1=0.51$ and $\lambda_2 = 0$ according to \cite{maes2009soft}, in order to obtain a model which is equivalent to a BSC with crossover probability $p=0.15$, which is often used in the PUF scenario. 
\cite{maes2009soft} also derived a cumulative density function and a probability density function of the error probability $P_{e}$ of an SRAM cell:
\begin{align*}
\cdf_{P_{e}}(x)= \cdf_{P_{one}}(x) + 1 - \cdf_{P_{one}}(1-x)
\end{align*}
\begin{equation*}
\pdf_{P_{e}}(x) =
\begin{cases}
\pdf_{P_{one}}(x) + \pdf_{P_{one}}(1-x)& \text{, if } x \textless 0.5, \\
0 & \text{, otherwise.}
\end{cases}
\label{eqn:pdfpe}
\end{equation*}

\section{Convolutional Codes}
\label{sec:convcodes}
Convolutional codes, introduced by Elias \cite{elias1955coding}, are widely used in applications.
Efficient methods for implementations in hardware make them suited for PUFs.
\cite{hiller2013breaking} applied convolutional codes to PUFs for the first time.
An optimized FPGA Viterbi decoder adapted to PUFs was designed in \cite{hiller2014seesaw}.
We explain encoding and decoding briefly, for a detailed description of convolutional codes we refer to \cite{costello2004error,johannesson1999fundamentals}.

\subsection{Description and Encoding}
\label{subsec:encoding}
Convolutional codes can be described using the parameters $(n,k,[\mu])$.
Using a code of rate $\frac{k}{n}$, we partition an information sequence into blocks of length $k$. 
Each information block is mapped to a codeword of length $n>k$.
In contrast to block codes, the codeword $\c_r$ depends on information block $\i_r$ and on the $\mu$ previous information blocks $\i_{r-1},\dots,\i_{r-\mu}$.	

\begin{figure}[htbp]
	\centering
	\includegraphics[width=0.4\linewidth]{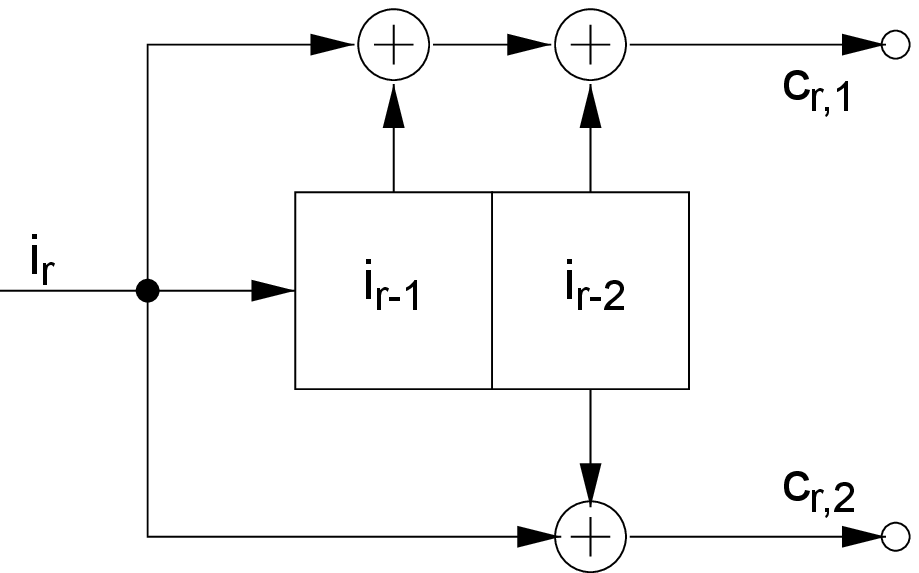}
	\caption{Encoder of a $(2,1,[2])$ convolutional code.}
	\label{fig:cc_encoder}
\end{figure}

Encoding of an information block can be done by calculating 
$\c_r = \i_r\G_0 + \i_{r-1}\G_1 + \dots \i_{r-\mu}\G_\mu$, or equivalently
$\c_r = (\i_r,\i_{r-1},\dots \i_{r-\mu}) \cdot (\G_0^T,\G_1^T,\dots,\G_\mu^T)^T$,
using $(k\times n)$ generator matrices $\G_0,\G_1,\dots,\G_\mu$. 
This process can be implemented by using a linear shift register for each of the $k$ inputs, $n$ outputs and $\mu$ memory elements per shift register, cf. Figure~\ref{fig:cc_encoder} for a $(2,1,[2])$ encoder with $\G_0 = (1,1)$, $\G_1 = (1,0)$, $\G_2 = (1,1)$.
Initially, all memory elements contain zeros. 
Information block $\i_r$ is mapped to codeword $\c_{r,1}\c_{r,2}$ by applying linear operations.
Then all bits move one memory element to the right. 
Bit $\i_r$ enters the leftmost memory element, bit $\i_{r-2}$ is dropped.
Encoding continues with block $\i_{r+1}$. 

\subsection{Viterbi Algorithm}
\begin{figure}[htbp]
	\centering
	\includegraphics[width=0.6\linewidth]{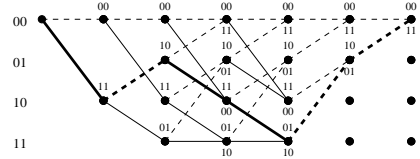}
	\caption{Trellis diagram for decoding a $(2,1,[2])$ code.}
	\label{fig:trellis2}
\end{figure}

The Viterbi algorithm realizes efficient \emph{maximum-likelihood decoding} \cite{viterbi1967error}.
We can visualize decoding using a graph structure called \emph{trellis} (cf. Figure~\ref{fig:trellis2} for the trellis of the $(2,1,[2])$ code introduced in Section~\ref{subsec:encoding}).
The nodes in the trellis represent the $2^{\mu}$ states of the shift register. 
Each node has two outgoing edges for encoding a 0 (dashed lines) or a 1 (solid lines).
Since the register is initialized with zeros, not all states occur in the beginning.
The edges are labeled with the corresponding codeword and represent the transitions to the next memory state.	
Hence, each path in the trellis represents a possible code sequence $\c$. The decoder aims to find the code sequence which was transmitted with highest probability when the sequence $\y$ was received, hence it maximizes $p(\y|\c)$.

Let $\c = (\c_1,\c_2,\dots)$ denote the code sequence which is sent over a noisy channel, $\c_i$ denotes a length-$n$ block.
Analog to this notation, $\y$ denotes the received sequence.
We define a \emph{submetric} and a \emph{metric}.

\begin{definition}
	For every edge, the submetric is defined as 
	\begin{align*}
	%\lambda_i = n - dist_{H}(\c_i,\y_i),
	\lambda_i = p(y_i|c_i) = 	   
	\begin{cases}
		p_i & \text{, } y_i \neq c_j \\
		1 - p_i & \text{, } y_i = c_i 
	\end{cases}.
	\end{align*}
	where $dist_{H}(\c_i,\y_i)$ denotes the Hamming distance of codeword $\c_i$ and received word $\y_i$.
	The metric of a node is the sum over all $\eta$ submetrics $\lambda_i$ on a path, i.e.
	$ \Lambda = \sum_{i=1}^{\eta}\lambda_i$ .
\end{definition}

Viterbi calculates the metric for all nodes in the trellis.
To calculate the metric for a specific node, for each incoming edge Viterbi adds the incoming submetric to the  metric of the corresponding node on the left side of the segment.
Viterbi stores the maximum reached metric for the node as its metric and adds the edge which contributes to this metric to its survivor path. 
To circumvent error propagation, often $m$ tail bits are inserted after $L$ information bits to reset the register (cf. Figure~\ref{fig:trellis2}, segments 5 and 6).
The bold path in Figure~\ref{fig:trellis2} is the survivor, which gives us $\c$ with $p(\y|\c)$ maximized.
The complexity of Viterbi increases exponentially in $\mu$ and $k$, and linearly in the number of info bits.

There is also a list decoding variant by \cite{schmidt2004finding}.
Using this modification of the Viterbi algorithm, each node stores the difference between the metric of the survivor and the second best path in addition to the metric.
When the maximum likelihood path was determined, the minimum of the	metric differences of all nodes on the ML path is selected, since the most unreliable decision was taken at the corresponding node.
At that node, we take the non-surviving edge to differ from the surviving path, and follow the second path backwards until it merges again with the survivor path.
By recursively applying this decision process, we can generate the $\mathcal{L}$ most likely code sequences.

\subsection{Sequential Decoding}
\label{subsec:sequential}

Viterbi's algorithm provides maximum-likelihood decoding.
However, its complexity grows exponentially when increasing $\mu$.
In this section we briefly discuss sequential decoding, introduced in \cite{wozencraft1957sequential}.
Since for this class of algorithms, complexity does not depend on the constraint length, larger constraint lengths can be used.
Sequential decoding is a almost maximum-likelihood decoding and can easily be implemented.
A further advantage in comparison to Viterbi is a runtime depending on the number of errors in the received sequence.
There exist different algorithms to realize sequential decoding, e.g. ZJ \cite{zigangirov1966some,jelinek1969fast} and Fano \cite{fano1963heuristic}.
We decided to use the Fano algorithm which is suitable for hardware implementations, since it requires essentially no memory \cite[Chapter~6]{johannesson1999fundamentals} and hence is a reasonable choice for the PUF scenario.
Compared to ZJ, the complexity of the Fano algorithm is slightly higher.
However, the higher complexity is only relevant in cases where many errors occur, since Fano backtracks more often than in cases with fewer errors.
For a small number of errors, ZJ is practically not faster, since it uses a stack which must be maintained during the decoding is performed. 
The Fano algorithm is said to be the most practical sequential decoding algorithm \cite[Section~6.10]{johannesson1999fundamentals}.

Sequential Decoding belongs to the class of backtracking algorithms which perform a depth first search in a tree.
The Fano algorithm traverses the code tree from the root until a leaf node is reached.
At each node, the algorithm can move either to one of its child nodes, or to its root.
The decision is based on the Fano metric: 
The bit metric for a BSC with crossover probability $p_i$ and rate $R$ is given by
\begin{align}
	M(y_i|c_i) &= \begin{cases}
		\log_2 2p_i - R &\text{ if $y_i \neq c_i$,} \\
		\log_2 2(1-p_i)-R& \text{ if $y_i =c_i$.}
	\end{cases}
	\label{eqn:fanometric}
\end{align}
The metric of a new node is calculated by adding the metric of the corresponding edge to the metric of the current node.

In each iteration, the algorithm examines which of the child nodes of the current node maximizes the metric (\ref{eqn:fanometric}), moves to that node and updates the metric.
This process continues, until the metric falls below a threshold $T$, what happens when it becomes unlikely that the correct path was chosen.
In that case, the next best path starting at the current node is examined. 
If all paths beginning at the current node are examined and the metric becomes lower than the threshold in all cases, the algorithm backtracks to the previous level. 
If the metric value of that preceding node also is below the threshold, the threshold is decreased by a threshold increment $\Delta$.
Every time a node is visited the first time, the threshold is increased by the maximum multiple of $\Delta$ such that the metric value is still larger than the threshold.

More information about sequential decoding with full length examples can be found in \cite[Chapter~13]{costello2004error}.

\section{Setup and Results}
\label{results}
In a first approach we extract $t$ bits from a PUF.
Using the model described in Section \ref{subsec:SRAMtheo}, each of these $t$ bits can be modeled as a BSC with crossover probability $p_i$ ($i = 1, . . . , t$). 
We select $s \leq t$ Bits $r_i$ ($i = 1, . . . , s$) which belong to BSCs with crossover probabilities $p_i < p_T$, where $p_T$ is a chosen threshold. 
Thus, unreliable PUF bits are ignored.

In a second step, we replace Viterbi by Fano algorithm, which is a sequential decoding approach, in order to gain performance for codes with large constraint lengths.

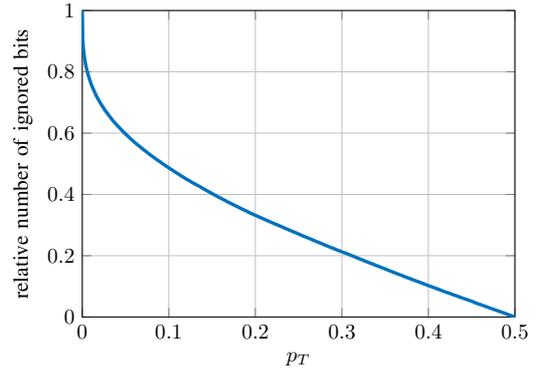
\begin{figure}
	\begin{center}
		\input{number_of_ignored_bits_plot.tex}
	\end{center}
	\caption{Relative number of ignored bits for a given threshold $p_T$}
	\label{fig:relative_number_of_ignored_bits}
\end{figure}

Figure~\ref{fig:relative_number_of_ignored_bits} shows the amount of wasted bits in our first approach.
In a third step we use all extracted bits without refusing the unreliable ones. 
Our goal is to study, whether the good channels compensate the bad ones when using soft information at the input of the decoder.
For a fair comparison we have to adjust the code rate for this scenario.

\subsection{Coding Theoretical Concepts}
Instead of using the SRAM reliability information only for choosing reliable bits from the PUF response, we can additionally translate them into soft information which can be used at the input of Viterbi's algorithm to improve the decoding results.
We calculate soft information
\begin{align}
s_{i}^{(j)} = (-1)^{h_{i} \oplus r_{i}^{(j)}} \cdot (\log(1-p_{e_{i}}) - \log(p_{e_{i}}) )
\label{eqn:si}
\end{align}
for cell $i$ at time instance $j$, where $r_{i}$ is the $i$-th response bit and $p_{e_{i}}$ is the error probability of SRAM cell $i$.
This formula is already given in \cite{maes2009soft}, however the authors use short linear block codes with soft decision decoding instead of convolutional codes. 
We obtain $p_{e_{i}}$ for all cells $i$ by using the model described in Section~\ref{subsec:SRAM}. In a practical scenario, the soft information can be obtained during the enrollment phase by doing multiple measurements for every individual PUF\cite{maes2009soft,maes2009low}. 
A technique how to obtain soft information from one single measurement was described in \cite{van2012soft}.

The simplified ROVA used in \cite{hiller2016}, has the drawback, that it calculates error probabilities for each symbol in order to compute the total error probability for the whole code sequence. 
As alternative to the calculation of error probabilities, we use a list decoding variant of the Viterbi algorithm according to \cite{schmidt2004finding}, which determines the $\mathcal{L}$ most reliable paths.
After computing a list, we can obtain reliability values by calculating 
\begin{align}
	\frac{p(r|c_i)}{p(r|c_j)}  = \frac{p(c_i|r)}{p(c_j|r)} \text{, where }p(c_i|r) = \prod_{i=1}^{n}q_i \\
	   \text{and }q_i =
	   \begin{cases}
	   	p_i & \text{, } r_i \neq c_i \\
	   	1 - p_i & \text{, } r_i = c_i 
	   \end{cases}.
\end{align}
for all candidate solutions $c_i,c_j$ and hence decide for the most likely code sequence.

Table~\ref{tab:codes} lists all convolutional codes used within this work.
Length of the used information sequence is 256 Bits with termination.

\begin{table}
	\begin{center}
		\caption{Convolutional codes used in this work.}
		\begin{tabular}{  c | c | c | r | r  | r }
			\multicolumn{3}{ c }{~} &\multicolumn{3}{ |c }{polynomials used (in octal)}\\ \hline
			$k$ &  $n$ & $\mu$& $g_2$ & $g_1$ & $g_0$\\ \hline
			1 & 2 & 2 & & 5 & 7 \\
			1 & 2 & 6 & & 133 & 171 \\
			1 & 2 & 7 & & 247 & 371 \\
			1 & 2 & 10&  & 3645 & 2671 \\
			1 & 2 & 14 & & 63057 & 44735 \\
			1 & 2 & 16 & & 313327 & 231721 \\
			1 & 3 & 6 & 133 & 165 & 171 \\
			1 & 3 & 7 & 225 & 331 & 367 \\
			1 & 3 & 8 & 557 & 663 & 711 \\
			1 & 3 & 9 & 1117 & 1365 & 1633 \\
			1 & 3 & 10 & 2353 & 2671 & 3175 
		\end{tabular} 
		\label{tab:codes}
	\end{center}
\end{table}

\subsection{Simulation Results}
\label{subsec:SRAMtheo}
The results presented in this section are based on the SRAM model introduced in~\cite{maes2009soft} and discussed in Section~\ref{subsec:SRAM}.
We perform simulations to compare our ideas to the results in~\cite{hiller2016}.
According to \cite{hiller2016}, we aim for an error probability in the order of $10^{-6}$.
For comparability we use $(2,1,[6])$ and $(2,1,[7])$ convolutional codes. 
Additionally, we use rate $\frac{1}{2}$ convolutional codes with memory length $10$, $14$, and $16$ to further improve the results.
For our experiments we used information sequences of length 256 and responses of length 512 + $\mu$. 
We want to emphasize, that in contrast to \cite{hiller2016} we do not use ROVA.

\subsubsection{Full Soft Information}
\label{subsec:expsoft}
In a first experiment, we use the same scenario as in \cite{hiller2016}, using thresholds $p_T\in\{0.3,0.2,0.1\}$. Instead of hard decision decoding, we use full soft information at the input of the Viterbi algorithm.

The results in Table~\ref{tab:exp1} show, that using soft information as input results in smaller error probabilities compared to \cite{hiller2016}.  
This becomes clear when comparing Column~4 and Column~7 in Table~\ref{tab:exp1}.
E.g., using a (2,1,[7]) code with $\pT=0.1$, \cite{hiller2016} achieved $P_{err} = 4.86\cdot10^{-3}$, using full soft information results in $P_{err} = 7.63\cdot10^{-4}$.

\subsubsection{Full Soft Information and List Decoding}
\label{subsec:expsoftlist}

%Applying list decoding further decreases the error probability. % (cf. column ``SD, $\mathcal{L}=3$'').

Applying list decoding allows to increase $\pT$ without loss.
This can be seen when studying for example the (2,1,[6]) code in Table~\ref{tab:exp1}.
In \cite{hiller2016} a key error probability of $1.08 \cdot 10^{-2}$ was obtained using $\pT = 0.1$. 
Using list decoding with list size $\mathcal{L}=4$ results in a key error probability of the same order, however we can increase $\pT$ to $0.2$.
Since more PUF bits can be used for the key, we can save $\frac{1}{3}$ of the extracted bits, as can be seen in Figure~\ref{fig:relative_number_of_ignored_bits}.

\begin{table*}[htbp]
	\centering
		\caption{Using rate $\frac{1}{2}$ codes. Comparison of  \cite{hiller2016} with hard decision decoding (HD) and soft decision decoding (SD), used with list size $\mathcal{L}$. Extr. means the number of simulated key extractions.$^*$ only 500.000 extractions were used in \cite{hiller2016}.} 
	\begin{tabular}{ | c | c | c | c | c | c |c | c | c |}
		\hline
		{\bf Code} & $\mathbf{p_T}$ & {\bf Extractions} & {\bf Ref.\cite{hiller2016}} & {\bf HD}, $\mathcal{L}=3$&{\bf HD}, $\mathcal{L}=4$ & {\bf SD}, $\mathcal{L}=1$ & {\bf SD}, $\mathcal{L}=3$ & {\bf SD}, $\mathcal{L}=4$ \\\hline
		%(2,1,[6]) & 0.3 & 500.000 & 4.11e-01 & 6.73e-01 &  6.67e-01 & 1.86e-01 & 1.30e-01 &1.2701e-01\\\hline
		(2,1,[6]) & 0.3 & 10.000.000 & 4.11e-01$^*$ & 6.72e-01 & 6.66e-01 & 1.86e-01 & 1.29e-01 &1.27e-01 \\ \hline
		%~ & 0.2 & 500.000 & 6.98e-02 & 1.83e-01 & 1.70e-01 & 3.61e-02 & 2.11e-02 &2.04e-02\\\hline
		~ & 0.2 & 10.000.000 & 6.98e-02$^*$ & 1.83e-01& 1.80e-01 & 3.62e-02 & 2.09e-02 &2.05e-02 \\\hline
		~ & 0.1 & 10.000.000 & $\leq$1.98e-02 & 1.04e-02& 1.02e-02 & 2.15e-03 & 1.09e-03 &1.07e-03\\\hline
		(2,1,[7]) & 0.3 &10.000.000 & -- & 4.98e-01 & 4.91e-01& 1.18e-01 &7.92e-02 & 7.74e-02\\\hline
		~ & 0.2 &10.000.000 & --  &8.37e-02 & 8.18e-02 & 1.80e-02 & 1.01e-02& 9.90e-03\\\hline
		~ & 0.1 &10.000.000 & $\leq$4.86e-03 & 2.68e-03& 2.62e-03 & 7.63e-04 & 3.80e-04& 3.73e-04\\\hline
		(2,1,[10]) & 0.3 &10.000.000 & -- & 3.05e-01  & 2.98e-01 & 2.97e-02 & 1.85e-02 & 1.79e-02\\\hline
		~ & 0.2 &10.000.000 & --  & 2.66e-02 & 2.57e-02 & 2.28e-03 & 1.21e-03 & 1.17e-03\\\hline
		~ & 0.1 &10.000.000 & -- & 3.20e-04 & 3.09e-04 & 3.90e-05 & 1.80e-05 & 1.80e-05\\\hline
		(2,1,[14]) & 0.3 &500.000 & --  & 1.20e-01 & 1.16e-01 & 4.17e-03 & 2.54e-03 &2.45e-03\\\hline
		~ & 0.2 &500.000 & --  & 3.37e-03& 3.24e-03 & 8.60e-05 & 4.20e-05 &4.00e-05\\\hline
		~ & 0.1 &500.000 & --  & 1.60e-05& 1.60e-05 &  {\bf  $<$2e-06}  &  {\bf  $<$2e-06} & {\bf  $<$2e-06}\\\hline
		(2,1,[16]) & 0.3 &500.000 & --  & 1.56e-03& 1.5e-03 & 2.00e-05 & {\bf $<$2e-06} &{\bf  $<$2e-06} \\\hline
		~ & 0.2 &500.000 & --  &  {\bf $<$2e-06}& 1.5e-03 & {\bf $<$2e-06}  &  {\bf  $<$2e-06} &{\bf $<$2e-06} \\\hline
		~ & 0.1 &500.000 & --  &  {\bf  $<$2e-06}& {\bf $<$2e-06}  & {\bf $<$2e-06}  &  {\bf  $<$2e-06} &{\bf  $<$2e-06} \\\hline
	\end{tabular}
	\label{tab:exp1}
\end{table*}

\subsubsection{Additional Use of Multiple Readouts}
\label{subsec:expmultiple}

The goal in this section is to further decrease the error probabilities gained in Section~\ref{subsec:expsoftlist}, by additionally using multiple readouts as done in \cite{hiller2016}.
Table~\ref{tab:exp2} presents the error probabilities obtained from simulations with list size $\mathcal{L}$ and using $m$ readouts.
Applying multiple readouts results in an essential improvement.
For example, considering a (2,1,[6]) code with $p = 0.1$, using full soft information with list size 3 and $m$ readouts is as good as using $m+1$ readouts in \cite{hiller2016}.
We can also use the same number of multiple readouts as in \cite{hiller2016} and do not use list decoding to result in the desired error probability $10^{-6}$. 
For convolutional codes with larger memory lengths, $p_T$ can be larger in order to achieve $P_{err} \approx 10^{-6}$.

\begin{figure}[htbp]
	\centering
	\includegraphics[width=0.8\linewidth]{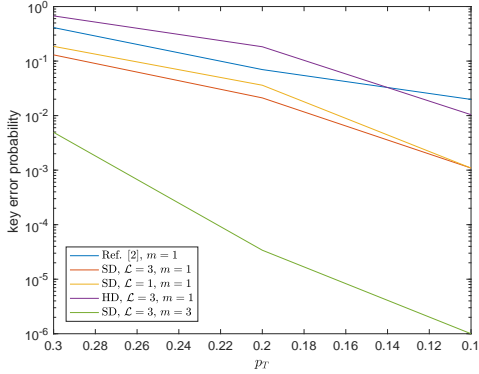}
	\caption{ $(2,1,\lbrack6\rbrack)$ code with soft information at the input, list decoding, and $m$ readouts.}
	\label{fig:plots}
\end{figure}

\begin{figure}[htbp]
	\centering
	\includegraphics[width=0.8\linewidth]{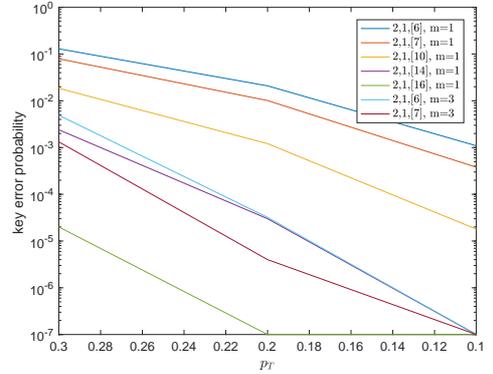}
	\caption{Comparison of convolutional codes with different memory lengths, soft input information, list decoding with list size $\mathcal{L}$, and $m$ readouts.}
	\label{fig:plots2}
\end{figure}

Figure~\ref{fig:plots} compares the key error probability of the (2,1,[6]) code from \cite{hiller2016} with our soft information input and list decoding approach for different values of $p_T$.
Note that the threshold measure used in \cite{hiller2016}, $p_{max}$, is specific for the \emph{Differential Sequence Coding (DSC)} scheme implemented in that reference, and hence is translated to $p_T$ in our tables.
Figure~\ref{fig:plots2} compares convolutional codes with different memory length, using soft information input, list size 3 and multiple readouts.
%Compared to \cite{hiller2016}, our approach also results in error probabilities smaller than $10^{-6}$ for $p_T=0.1$.

\begin{table*}[btp]
	\centering
		\caption{Adding the concept of using $m$ readouts to the techniques of soft information and list decoding. Using rate $\frac{1}{2}$ codes.}
	\begin{tabular}{ | c | c | c|  c | c | c | c |c | c |}
		\hline
		{\bf Code} & $\mathbf{p_T}$ & {\bf Extractions} & {\bf Ref.\cite{hiller2016}} & {\bf Ref.\cite{hiller2016}} & {\bf HD}, $\mathcal{L}=3$  &{\bf SD}, $\mathcal{L}=1$, & {\bf SD}, $\mathcal{L}=3$, & {\bf SD}, $\mathcal{L}=3$, \\
		~ & ~ & ~ & $m=1$ & $m=3$& $m=3$& $m=3$ &$m=2$ & $m=3$ \\ \hline
		(2,1,[6]) & 0.3 & 500.000 & 4.11e-01 &-- & 6.32e-01 &  7.81e-03 & 3.58e-02 & 4.93e-03 \\\hline
		~ & ~ & 10.000.000 & -- &-- & 6.32e-01 & 7.86e-03 & 3.59e-02 & 4.91e-03 \\ \hline
		~ & 0.2 & 500.000 & 4.11e-01 & --& 1.84e-01 & 5.40e-05 & 1.35e-03 & 3.40e-05\\\hline
		~ & ~ & 10.000.000 & -- &-- & 1.84e-01 & 6.10e-05 & 1.36e-03 & 3.20e-05 \\ \hline
		~ & 0.1 & 10.000.000 &  $\leq$1.98e-02& {\bf7.6e-07} & 3.27e-02 &  {\bf $<$1e-07}  & {\bf 6.00e-06} &  {\bf $<$1e-07} \\\hline
		(2,1,[7]) & 0.3 & 10.000.000 & --& -- &4.89e-01 &  2.19e-03 & 1.49e-02 & 1.32e-03\\\hline
		~ & 0.2 & 10.000.000 &  --~& -- & 1.05e-01 & {\bf 8.00e-06}  & 3.44e-04 & {\bf 4.00e-06}\\\hline
		~ & 0.1 & 10.000.000 & $\leq$4.86e-03& {\bf$<$1e-07}& 2.66e-02 &  {\bf $<$1e-07}  & {\bf 1.00e-06} &  {\bf $<$1e-07}\\\hline
		(2,1,[10]) & 0.3 & 10.000.000 & -- &-- & 3.25e-01 &  4.70e-05 & 1.02e-03 & 2.85e-04\\\hline
		~ & 0.2 & 10.000.000 & -- & -- & 6.43e-02 &  {\bf $<$1e-07}  & {\bf 5.00e-06} &  {\bf $<$1e-07}\\\hline
		~ & 0.1 & 10.000.000 & --& --& 3.98e-02 &  {\bf $<$1e-07}  & {\bf $<$1e-07} &  {\bf $<$1e-07}\\\hline
	\end{tabular}
	\label{tab:exp2}
\end{table*}		

\subsubsection{Sequential Decoding}
Our results using the Viterbi algorithm show that the error probability depends on the memory length, which is a well-known fact in coding theory.
As indicated by the tables, to reach an error probability of $10^{-6}$, a comparably large memory length is needed.
However, the complexity of the Viterbi decoder depends exponentially on the memory and becomes infeasible for $\mu \approx 20$ in case of $k=1$ and even lower values for $k>1$.
This drawback can be circumvented by using sequential decoding as discussed in Section~\ref{subsec:sequential}.
Table~\ref{tab:exp_sequential} shows our results.

\begin{table}[hbtp]
	\begin{center}
		\caption{Performance of sequential decoding using Fano's Algorithm with soft information input, $\mathcal{L}=1$, and $p_T = 0.3$} 
		\begin{tabular}{ | c | c | c |}
			\hline
			{\bf Code} &  {\bf Extractions} & {\bf SD\footnotemark}  \\ \hline
			(2,1,[16]) & 10.000.000 & $\sim10^{-5}$    \\\hline	
			(2,1,[20]) & 10.000.000 & $\sim10^{-6}$   \\\hline	
			(2,1,[24]) & 10.000.000 & $\sim10^{-7}$    \\\hline	
		\end{tabular}
		\label{tab:exp_sequential}
	\end{center}
\end{table}
\footnotetext{Due to time restrictions, the numbers in the table are estimations based on simulations with similar parameters. Simulations will have finished until the final deadline and numbers in the table will be replaced.}

\subsubsection{Without Refusing Unreliable Bits}
\label{subesc:expnodsc}

\begin{table*}
	\centering
	\caption{Applying a rate $\frac{1}{3}$ convolutional code, using all response bits, even the unreliable ones. The parameters $\mu,\mathcal{L},m$ denote memory length, list size and number of readouts respectively.} 
	\begin{center}
		\begin{tabular}{ | c | c | c | c | c | c | c |}
			\hline
			{\bf Code} &  {\bf Extractions} & {\bf SD},  $\mathcal{L}=1$ & {\bf HD}, $\mathcal{L}=3$& {\bf SD}, $\mathcal{L}=3$ &{\bf SD}, $\mathcal{L}=3$, $m=3$  \\ \hline
			(3,1,[6]) &10.000.000 & 3.98e-02 & 6.59e-01 & 2.24e-02 & 5.70e-05 \\\hline
			(3,1,[7]) &10.000.000 & 1.73e-02 & 5.93e-01 &  9.43e-03& {\bf 6.00e-06} \\\hline
			(3,1,[8]) &10.000.000 & 9.72e-03 & 5.09e-01& 5.14e-03& {\bf 1.00e-06}  \\\hline
			(3,1,[9]) &10.000.000 & 5.07e-03 & 4.28e-01 & 2.65e-03&  {\bf $<$1e-07} \\\hline
			(3,1,[10]) &10.000.000 & 2.30e-03 & 3.39e-01 & 1.17e-03&  {\bf $<$1e-07} \\\hline	
		\end{tabular}
		\label{tab:exp3_2}
	\end{center}
\end{table*}

A drawback of the approaches described so far is, that not all response bits are used.
Hence more bits than needed have to be extracted from the PUF, what can be time- and area-consuming (cf. Figure~\ref{fig:relative_number_of_ignored_bits}).
Instead of refusing response bits with an error probability higher than a certain threshold $p_T$, we can use all response bits and use their reliability information as soft input to a Viterbi decoder.
The idea is, that the soft information of reliable bits compensates unreliable bits.
As seen in Figure~\ref{fig:relative_number_of_ignored_bits}, using $\pT \approx 0.2$ results in $\approx\frac{1}{3}$ of ignored bits. 
For a fair comparison with \cite{hiller2016}, we therefore use rate $\frac{1}{3}$ instead of rate $\frac{1}{2}$ codes.

The results in Table~\ref{tab:exp3_2} show that this method significantly improves the previous approaches.
This is obvious by comparing the columns ``SD, $\mathcal{L} = 1$'' in Tables~\ref{tab:exp1}~and~\ref{tab:exp3_2}.
For example we can realize, that the results for (2,1,[6]), (2,1,[7]), and (2,1,[10]) for $\pT = 0.2$ in Table~\ref{tab:exp1} coincide with the corresponding results in Table~\ref{tab:exp3_2}.
We want to emphasize, that this implies advantages to the previous results.
First, we do not ignore extracted PUF bits, second, ROVA can be avoided.

\section{Conclusion}
This work studied how previous approaches that use convolutional codes for PUFs can be improved by using soft information, list decoding and sequential decoding.
Our results show that the selection of reliable bits from a PUF response is not needed.
Reliability values of the individual response bits are translated into soft information which is used at the input of Viterbi.
Using these soft information, reliable channels compensate unreliable ones.
A rate $\frac{1}{3}$ convolutional code of memory length 7 is sufficient in order to result in a key error probability of order $10^{-6}$.
The obtained error probabilities can further be decreased by increasing memory length or using a larger list size for list decoding.
Using a convolutional code with more than 8 memory elements, the error probability becomes smaller than $10^{-7}$.
Using sequential decoding decreases decoding complexity when using codes with large memory length.
To summarize, our strategy has two advantages compared to previous approaches.
First, all response bits are used.
Second, we avoid overhead since our work does not require ROVA.

%Our test with practical data even indicate that assuming a BSC with crossover probability $p \approx 0.15$ seems to be to pessimistic when considering specific PUF implementations.
%We observed, that we can chose a much lower $\mu$ then assumed by theoretical models and hence in many practical cases multiple readouts and list decoding are not even needed.

%\section*{Acknowledgment}
%Our implementation of the experiments is based on a convolutional code library implemented by Walter Schnug and a Fano algorithm implemented by our student Liming Fan.

\bibliographystyle{IEEEtran}
\bibliography{main}


\end{document}

%% file: preamble.tex
\usepackage[latin9]{inputenc}
\usepackage[T1]{fontenc}
\usepackage{url}
\usepackage{ifthen}
\usepackage[cmex10]{amsmath} % Use the [cmex10] option 
\usepackage{pgfplots}
\pgfplotsset{compat=newest}
\usepackage{tikz}
\usepackage{multirow}
\usepackage{amsmath}
\usepackage{amssymb}
\usepackage{amsthm}
\usepackage{mathrsfs}
\usepackage{graphicx}
\usepackage{color}
\usepackage[linesnumbered,ruled,vlined,titlenumbered]{algorithm2e}
\usepackage{flushend}

\renewcommand{\i}{\vec{i}}
\renewcommand{\c}{\vec{c}}
\newcommand{\y}{\vec{y}}

\renewcommand{\pi}{p_i}
\newcommand{\pT}{p_T}

\newcommand{\G}{\vec{G}}

\newcommand{\cdf}{\mathbf{cdf}}
\newcommand{\pdf}{\mathbf{pdf}}

\newtheorem{theorem}{Theorem}

\newtheorem{definition}[theorem]{Definition}

\usepackage{varioref}        % for \vref{:...} gives "As 1.2 on page 4"
\usepackage{fancyref}

\makeatletter
\def\mkfancyprefix#1#2{%
\expandafter\def\csname fancyref#1labelprefix\endcsname{#1}%
\begingroup\def\x{\endgroup\frefformat{plain}}%
    \expandafter\x\csname fancyref#1labelprefix\endcsname
    {\MakeLowercase{#2}\fancyrefdefaultspacing##1}%
\begingroup\def\x{\endgroup\Frefformat{plain}}%
    \expandafter\x\csname fancyref#1labelprefix\endcsname
    {#2\fancyrefdefaultspacing##1}%
\begingroup\def\x{\endgroup\frefformat{vario}}%
    \expandafter\x\csname fancyref#1labelprefix\endcsname
    {\MakeLowercase{#2}\fancyrefdefaultspacing##1##3}%
\begingroup\def\x{\endgroup\Frefformat{vario}}%
    \expandafter\x\csname fancyref#1labelprefix\endcsname
    {#2\fancyrefdefaultspacing##1##3}%
}
\makeatother
\fancyrefchangeprefix{\fancyrefeqlabelprefix}{eqn}
\mkfancyprefix{ssec}{Section}
\mkfancyprefix{tbl}{Table}
\mkfancyprefix{thm}{Theorem}
\mkfancyprefix{lem}{Lemma}
\mkfancyprefix{cor}{Corollary}
\mkfancyprefix{prop}{Proposition}
\mkfancyprefix{prob}{Problem}
\mkfancyprefix{alg}{Algorithm}
\mkfancyprefix{inv}{Invariant}
\mkfancyprefix{ex}{Example}
\mkfancyprefix{line}{Line}
\mkfancyprefix{def}{Definition}
\mkfancyprefix{itm}{Item}
\mkfancyprefix{app}{Appendix}
\mkfancyprefix{rem}{Remark}
\newcommand{\cref}[1]{\Fref{#1}}

\makeatletter
\newcommand{\removelatexerror}{\let\@latex@error\@gobble}
\makeatother

\renewcommand{\vec}[1]{\ensuremath{\mathbf{#1}}}

\renewcommand{\c}{\vec{c}}

%% file: number_of_ignored_bits_plot.tex
\definecolor{mycolor1}{rgb}{0.00000,0.44700,0.74100}%
\begin{tikzpicture}[scale=0.80]

\begin{axis}[%
width=2.8in,
height=2in,
at={(0.758in,0.481in)},
scale only axis,
xmin=0,
xmax=0.5,
xlabel={$p_T$},
xmajorgrids,
ymin=0,
ymax=1,
ylabel={relative number of ignored bits},
ymajorgrids,
axis background/.style={fill=white}
]
\addplot [color=mycolor1,solid, ultra thick,forget plot]
  table[row sep=crcr]{%
0	1\\
0.0005005005005005	0.90597\\
0.001001001001001	0.88423\\
0.0015015015015015	0.86899\\
0.002002002002002	0.85726\\
0.0025025025025025	0.84686\\
0.003003003003003	0.8376\\
0.0035035035035035	0.82979\\
0.004004004004004	0.82232\\
0.0045045045045045	0.81547\\
0.005005005005005	0.80949\\
0.00550550550550551	0.80358\\
0.00600600600600601	0.79811\\
0.00650650650650651	0.79323\\
0.00700700700700701	0.78829\\
0.00750750750750751	0.78345\\
0.00800800800800801	0.77892\\
0.00850850850850851	0.77464\\
0.00900900900900901	0.77068\\
0.00950950950950951	0.76643\\
0.01001001001001	0.76227\\
0.0105105105105105	0.7584\\
0.011011011011011	0.75474\\
0.0115115115115115	0.75105\\
0.012012012012012	0.74772\\
0.0125125125125125	0.74444\\
0.013013013013013	0.74147\\
0.0135135135135135	0.73845\\
0.014014014014014	0.73514\\
0.0145145145145145	0.73202\\
0.015015015015015	0.729\\
0.0155155155155155	0.72597\\
0.016016016016016	0.72295\\
0.0165165165165165	0.72019\\
0.017017017017017	0.71732\\
0.0175175175175175	0.71487\\
0.018018018018018	0.71236\\
0.0185185185185185	0.70967\\
0.019019019019019	0.70715\\
0.0195195195195195	0.70474\\
0.02002002002002	0.70236\\
0.0205205205205205	0.69991\\
0.021021021021021	0.69766\\
0.0215215215215215	0.69541\\
0.022022022022022	0.69301\\
0.0225225225225225	0.69079\\
0.023023023023023	0.68829\\
0.0235235235235235	0.68619\\
0.024024024024024	0.68421\\
0.0245245245245245	0.68193\\
0.025025025025025	0.67985\\
0.0255255255255255	0.67759\\
0.026026026026026	0.67563\\
0.0265265265265265	0.67356\\
0.027027027027027	0.67145\\
0.0275275275275275	0.66964\\
0.028028028028028	0.6677\\
0.0285285285285285	0.66628\\
0.029029029029029	0.6644\\
0.0295295295295295	0.66223\\
0.03003003003003	0.66022\\
0.0305305305305305	0.65845\\
0.031031031031031	0.65644\\
0.0315315315315315	0.65459\\
0.032032032032032	0.65285\\
0.0325325325325325	0.65098\\
0.033033033033033	0.64924\\
0.0335335335335335	0.64742\\
0.034034034034034	0.64574\\
0.0345345345345345	0.64394\\
0.035035035035035	0.642\\
0.0355355355355355	0.64023\\
0.036036036036036	0.6387\\
0.0365365365365365	0.63706\\
0.037037037037037	0.63532\\
0.0375375375375375	0.63362\\
0.038038038038038	0.63223\\
0.0385385385385385	0.63077\\
0.039039039039039	0.62917\\
0.0395395395395395	0.62765\\
0.04004004004004	0.62621\\
0.0405405405405405	0.62453\\
0.041041041041041	0.62309\\
0.0415415415415415	0.62153\\
0.042042042042042	0.62008\\
0.0425425425425425	0.6187\\
0.043043043043043	0.61706\\
0.0435435435435435	0.61557\\
0.044044044044044	0.61408\\
0.0445445445445445	0.61256\\
0.045045045045045	0.61112\\
0.0455455455455455	0.60973\\
0.046046046046046	0.60815\\
0.0465465465465465	0.60679\\
0.047047047047047	0.60527\\
0.0475475475475475	0.60379\\
0.048048048048048	0.60256\\
0.0485485485485486	0.60119\\
0.049049049049049	0.59993\\
0.0495495495495495	0.5984\\
0.0500500500500501	0.59723\\
0.0505505505505505	0.59571\\
0.0510510510510511	0.59432\\
0.0515515515515515	0.59305\\
0.0520520520520521	0.59169\\
0.0525525525525526	0.59028\\
0.0530530530530531	0.58898\\
0.0535535535535536	0.58768\\
0.0540540540540541	0.58635\\
0.0545545545545546	0.58496\\
0.0550550550550551	0.58377\\
0.0555555555555556	0.5824\\
0.0560560560560561	0.581\\
0.0565565565565566	0.57991\\
0.0570570570570571	0.57869\\
0.0575575575575576	0.57737\\
0.0580580580580581	0.57599\\
0.0585585585585586	0.5746\\
0.0590590590590591	0.57332\\
0.0595595595595596	0.57217\\
0.0600600600600601	0.57081\\
0.0605605605605606	0.56964\\
0.0610610610610611	0.56834\\
0.0615615615615616	0.56731\\
0.0620620620620621	0.56611\\
0.0625625625625626	0.56504\\
0.0630630630630631	0.56387\\
0.0635635635635636	0.56268\\
0.0640640640640641	0.56152\\
0.0645645645645646	0.56056\\
0.0650650650650651	0.55937\\
0.0655655655655656	0.5581\\
0.0660660660660661	0.55689\\
0.0665665665665666	0.55575\\
0.0670670670670671	0.55448\\
0.0675675675675676	0.5533\\
0.0680680680680681	0.55211\\
0.0685685685685686	0.55094\\
0.0690690690690691	0.54996\\
0.0695695695695696	0.54896\\
0.0700700700700701	0.54787\\
0.0705705705705706	0.5467\\
0.0710710710710711	0.5456\\
0.0715715715715716	0.54446\\
0.0720720720720721	0.54324\\
0.0725725725725726	0.54227\\
0.0730730730730731	0.54113\\
0.0735735735735736	0.53997\\
0.0740740740740741	0.53904\\
0.0745745745745746	0.53792\\
0.0750750750750751	0.53673\\
0.0755755755755756	0.53555\\
0.0760760760760761	0.5345\\
0.0765765765765766	0.53343\\
0.0770770770770771	0.5323\\
0.0775775775775776	0.53121\\
0.0780780780780781	0.53028\\
0.0785785785785786	0.5291\\
0.0790790790790791	0.52814\\
0.0795795795795796	0.52709\\
0.0800800800800801	0.52601\\
0.0805805805805806	0.52496\\
0.0810810810810811	0.52388\\
0.0815815815815816	0.52285\\
0.0820820820820821	0.52184\\
0.0825825825825826	0.52071\\
0.0830830830830831	0.5197\\
0.0835835835835836	0.51873\\
0.0840840840840841	0.51763\\
0.0845845845845846	0.51648\\
0.0850850850850851	0.51537\\
0.0855855855855856	0.51449\\
0.0860860860860861	0.51349\\
0.0865865865865866	0.5126\\
0.0870870870870871	0.51164\\
0.0875875875875876	0.51067\\
0.0880880880880881	0.50968\\
0.0885885885885886	0.50887\\
0.0890890890890891	0.50788\\
0.0895895895895896	0.5069\\
0.0900900900900901	0.50594\\
0.0905905905905906	0.50503\\
0.0910910910910911	0.50407\\
0.0915915915915916	0.50288\\
0.0920920920920921	0.50176\\
0.0925925925925926	0.5007\\
0.0930930930930931	0.49984\\
0.0935935935935936	0.49878\\
0.0940940940940941	0.49767\\
0.0945945945945946	0.49681\\
0.0950950950950951	0.49584\\
0.0955955955955956	0.49488\\
0.0960960960960961	0.49402\\
0.0965965965965966	0.49298\\
0.0970970970970971	0.49202\\
0.0975975975975976	0.49127\\
0.0980980980980981	0.49041\\
0.0985985985985986	0.48931\\
0.0990990990990991	0.48838\\
0.0995995995995996	0.48733\\
0.1001001001001	0.48625\\
0.100600600600601	0.48539\\
0.101101101101101	0.4845\\
0.101601601601602	0.48357\\
0.102102102102102	0.48282\\
0.102602602602603	0.48198\\
0.103103103103103	0.48093\\
0.103603603603604	0.48022\\
0.104104104104104	0.47936\\
0.104604604604605	0.47835\\
0.105105105105105	0.47722\\
0.105605605605606	0.47617\\
0.106106106106106	0.47523\\
0.106606606606607	0.47433\\
0.107107107107107	0.4733\\
0.107607607607608	0.47258\\
0.108108108108108	0.47177\\
0.108608608608609	0.47078\\
0.109109109109109	0.4699\\
0.10960960960961	0.469\\
0.11011011011011	0.46824\\
0.110610610610611	0.46729\\
0.111111111111111	0.46633\\
0.111611611611612	0.4654\\
0.112112112112112	0.46452\\
0.112612612612613	0.46382\\
0.113113113113113	0.46266\\
0.113613613613614	0.46182\\
0.114114114114114	0.461\\
0.114614614614615	0.46014\\
0.115115115115115	0.45921\\
0.115615615615616	0.45822\\
0.116116116116116	0.45738\\
0.116616616616617	0.45659\\
0.117117117117117	0.45569\\
0.117617617617618	0.45477\\
0.118118118118118	0.45389\\
0.118618618618619	0.45307\\
0.119119119119119	0.45205\\
0.11961961961962	0.4514\\
0.12012012012012	0.45031\\
0.120620620620621	0.44947\\
0.121121121121121	0.44863\\
0.121621621621622	0.44784\\
0.122122122122122	0.44707\\
0.122622622622623	0.44621\\
0.123123123123123	0.44542\\
0.123623623623624	0.4447\\
0.124124124124124	0.44391\\
0.124624624624625	0.44313\\
0.125125125125125	0.44243\\
0.125625625625626	0.44155\\
0.126126126126126	0.44063\\
0.126626626626627	0.4397\\
0.127127127127127	0.4388\\
0.127627627627628	0.43792\\
0.128128128128128	0.43717\\
0.128628628628629	0.43635\\
0.129129129129129	0.43566\\
0.12962962962963	0.43483\\
0.13013013013013	0.4342\\
0.130630630630631	0.43337\\
0.131131131131131	0.43265\\
0.131631631631632	0.43181\\
0.132132132132132	0.43096\\
0.132632632632633	0.43039\\
0.133133133133133	0.42966\\
0.133633633633634	0.42891\\
0.134134134134134	0.42811\\
0.134634634634635	0.42726\\
0.135135135135135	0.42651\\
0.135635635635636	0.42575\\
0.136136136136136	0.42495\\
0.136636636636637	0.42414\\
0.137137137137137	0.42338\\
0.137637637637638	0.42249\\
0.138138138138138	0.42175\\
0.138638638638639	0.42077\\
0.139139139139139	0.42004\\
0.13963963963964	0.41928\\
0.14014014014014	0.41837\\
0.140640640640641	0.41747\\
0.141141141141141	0.41671\\
0.141641641641642	0.41598\\
0.142142142142142	0.41516\\
0.142642642642643	0.41438\\
0.143143143143143	0.41351\\
0.143643643643644	0.4128\\
0.144144144144144	0.41209\\
0.144644644644645	0.41126\\
0.145145145145145	0.41039\\
0.145645645645646	0.40957\\
0.146146146146146	0.40886\\
0.146646646646647	0.40809\\
0.147147147147147	0.40742\\
0.147647647647648	0.40649\\
0.148148148148148	0.40568\\
0.148648648648649	0.40487\\
0.149149149149149	0.40409\\
0.14964964964965	0.40338\\
0.15015015015015	0.40256\\
0.150650650650651	0.40183\\
0.151151151151151	0.40099\\
0.151651651651652	0.4002\\
0.152152152152152	0.39928\\
0.152652652652653	0.39853\\
0.153153153153153	0.39772\\
0.153653653653654	0.39709\\
0.154154154154154	0.39639\\
0.154654654654655	0.39572\\
0.155155155155155	0.39494\\
0.155655655655656	0.39421\\
0.156156156156156	0.39332\\
0.156656656656657	0.39255\\
0.157157157157157	0.39183\\
0.157657657657658	0.39101\\
0.158158158158158	0.39022\\
0.158658658658659	0.38945\\
0.159159159159159	0.38894\\
0.15965965965966	0.38817\\
0.16016016016016	0.38744\\
0.160660660660661	0.38676\\
0.161161161161161	0.38615\\
0.161661661661662	0.3853\\
0.162162162162162	0.38457\\
0.162662662662663	0.38385\\
0.163163163163163	0.38315\\
0.163663663663664	0.38248\\
0.164164164164164	0.38174\\
0.164664664664665	0.38103\\
0.165165165165165	0.38013\\
0.165665665665666	0.37943\\
0.166166166166166	0.37859\\
0.166666666666667	0.37788\\
0.167167167167167	0.377\\
0.167667667667668	0.37618\\
0.168168168168168	0.37539\\
0.168668668668669	0.37473\\
0.169169169169169	0.37394\\
0.16966966966967	0.37329\\
0.17017017017017	0.37266\\
0.170670670670671	0.37202\\
0.171171171171171	0.37137\\
0.171671671671672	0.37065\\
0.172172172172172	0.36988\\
0.172672672672673	0.36917\\
0.173173173173173	0.36853\\
0.173673673673674	0.36784\\
0.174174174174174	0.36698\\
0.174674674674675	0.36629\\
0.175175175175175	0.36555\\
0.175675675675676	0.36492\\
0.176176176176176	0.36423\\
0.176676676676677	0.36352\\
0.177177177177177	0.36284\\
0.177677677677678	0.36223\\
0.178178178178178	0.36147\\
0.178678678678679	0.36086\\
0.179179179179179	0.36017\\
0.17967967967968	0.35944\\
0.18018018018018	0.35876\\
0.180680680680681	0.3581\\
0.181181181181181	0.35735\\
0.181681681681682	0.35662\\
0.182182182182182	0.35589\\
0.182682682682683	0.35508\\
0.183183183183183	0.35435\\
0.183683683683684	0.35374\\
0.184184184184184	0.35308\\
0.184684684684685	0.35224\\
0.185185185185185	0.35157\\
0.185685685685686	0.35079\\
0.186186186186186	0.35015\\
0.186686686686687	0.3495\\
0.187187187187187	0.34894\\
0.187687687687688	0.34838\\
0.188188188188188	0.34771\\
0.188688688688689	0.34707\\
0.189189189189189	0.34637\\
0.18968968968969	0.34571\\
0.19019019019019	0.34487\\
0.190690690690691	0.3441\\
0.191191191191191	0.34339\\
0.191691691691692	0.34281\\
0.192192192192192	0.34201\\
0.192692692692693	0.34134\\
0.193193193193193	0.34057\\
0.193693693693694	0.3399\\
0.194194194194194	0.33934\\
0.194694694694695	0.33877\\
0.195195195195195	0.33809\\
0.195695695695696	0.3374\\
0.196196196196196	0.33675\\
0.196696696696697	0.33603\\
0.197197197197197	0.33543\\
0.197697697697698	0.33474\\
0.198198198198198	0.33408\\
0.198698698698699	0.33334\\
0.199199199199199	0.33268\\
0.1996996996997	0.33205\\
0.2002002002002	0.33139\\
0.200700700700701	0.33092\\
0.201201201201201	0.33041\\
0.201701701701702	0.3298\\
0.202202202202202	0.32918\\
0.202702702702703	0.32842\\
0.203203203203203	0.32784\\
0.203703703703704	0.32716\\
0.204204204204204	0.32652\\
0.204704704704705	0.32592\\
0.205205205205205	0.32526\\
0.205705705705706	0.32468\\
0.206206206206206	0.32397\\
0.206706706706707	0.32345\\
0.207207207207207	0.32295\\
0.207707707707708	0.32245\\
0.208208208208208	0.32185\\
0.208708708708709	0.32107\\
0.209209209209209	0.32043\\
0.20970970970971	0.31977\\
0.21021021021021	0.31924\\
0.210710710710711	0.31851\\
0.211211211211211	0.31791\\
0.211711711711712	0.31725\\
0.212212212212212	0.31665\\
0.212712712712713	0.31615\\
0.213213213213213	0.31544\\
0.213713713713714	0.31479\\
0.214214214214214	0.31414\\
0.214714714714715	0.31345\\
0.215215215215215	0.31287\\
0.215715715715716	0.3123\\
0.216216216216216	0.31177\\
0.216716716716717	0.3112\\
0.217217217217217	0.31066\\
0.217717717717718	0.31\\
0.218218218218218	0.3093\\
0.218718718718719	0.30853\\
0.219219219219219	0.30796\\
0.21971971971972	0.30728\\
0.22022022022022	0.30666\\
0.220720720720721	0.30608\\
0.221221221221221	0.3056\\
0.221721721721722	0.30497\\
0.222222222222222	0.3043\\
0.222722722722723	0.30364\\
0.223223223223223	0.303\\
0.223723723723724	0.30247\\
0.224224224224224	0.30195\\
0.224724724724725	0.30138\\
0.225225225225225	0.30072\\
0.225725725725726	0.30011\\
0.226226226226226	0.29956\\
0.226726726726727	0.29903\\
0.227227227227227	0.29853\\
0.227727727727728	0.29785\\
0.228228228228228	0.29724\\
0.228728728728729	0.29665\\
0.229229229229229	0.29602\\
0.22972972972973	0.29547\\
0.23023023023023	0.29494\\
0.230730730730731	0.29423\\
0.231231231231231	0.29357\\
0.231731731731732	0.29291\\
0.232232232232232	0.29227\\
0.232732732732733	0.29155\\
0.233233233233233	0.29092\\
0.233733733733734	0.29028\\
0.234234234234234	0.2897\\
0.234734734734735	0.28919\\
0.235235235235235	0.28857\\
0.235735735735736	0.28793\\
0.236236236236236	0.28728\\
0.236736736736737	0.28679\\
0.237237237237237	0.28629\\
0.237737737737738	0.28568\\
0.238238238238238	0.28503\\
0.238738738738739	0.28432\\
0.239239239239239	0.28366\\
0.23973973973974	0.28315\\
0.24024024024024	0.28252\\
0.240740740740741	0.28197\\
0.241241241241241	0.28141\\
0.241741741741742	0.28087\\
0.242242242242242	0.28032\\
0.242742742742743	0.27975\\
0.243243243243243	0.27909\\
0.243743743743744	0.27843\\
0.244244244244244	0.27781\\
0.244744744744745	0.27726\\
0.245245245245245	0.27664\\
0.245745745745746	0.276\\
0.246246246246246	0.27544\\
0.246746746746747	0.275\\
0.247247247247247	0.27435\\
0.247747747747748	0.2735\\
0.248248248248248	0.27285\\
0.248748748748749	0.27228\\
0.249249249249249	0.27168\\
0.24974974974975	0.27114\\
0.25025025025025	0.2706\\
0.250750750750751	0.26993\\
0.251251251251251	0.26934\\
0.251751751751752	0.26866\\
0.252252252252252	0.26801\\
0.252752752752753	0.26741\\
0.253253253253253	0.26676\\
0.253753753753754	0.266\\
0.254254254254254	0.2655\\
0.254754754754755	0.26494\\
0.255255255255255	0.26438\\
0.255755755755756	0.26377\\
0.256256256256256	0.26327\\
0.256756756756757	0.26265\\
0.257257257257257	0.26214\\
0.257757757757758	0.2616\\
0.258258258258258	0.26096\\
0.258758758758759	0.26037\\
0.259259259259259	0.25994\\
0.25975975975976	0.25928\\
0.26026026026026	0.25879\\
0.260760760760761	0.25812\\
0.261261261261261	0.25747\\
0.261761761761762	0.25697\\
0.262262262262262	0.25637\\
0.262762762762763	0.2557\\
0.263263263263263	0.25514\\
0.263763763763764	0.25457\\
0.264264264264264	0.25404\\
0.264764764764765	0.2534\\
0.265265265265265	0.25292\\
0.265765765765766	0.2522\\
0.266266266266266	0.25165\\
0.266766766766767	0.25104\\
0.267267267267267	0.25021\\
0.267767767767768	0.24953\\
0.268268268268268	0.24895\\
0.268768768768769	0.24841\\
0.269269269269269	0.2478\\
0.26976976976977	0.24707\\
0.27027027027027	0.24648\\
0.270770770770771	0.24597\\
0.271271271271271	0.24543\\
0.271771771771772	0.24501\\
0.272272272272272	0.24455\\
0.272772772772773	0.24401\\
0.273273273273273	0.24344\\
0.273773773773774	0.24281\\
0.274274274274274	0.24226\\
0.274774774774775	0.24155\\
0.275275275275275	0.24085\\
0.275775775775776	0.24031\\
0.276276276276276	0.23976\\
0.276776776776777	0.23921\\
0.277277277277277	0.23845\\
0.277777777777778	0.23788\\
0.278278278278278	0.23744\\
0.278778778778779	0.23676\\
0.279279279279279	0.23605\\
0.27977977977978	0.23544\\
0.28028028028028	0.23486\\
0.280780780780781	0.23434\\
0.281281281281281	0.23379\\
0.281781781781782	0.23327\\
0.282282282282282	0.23256\\
0.282782782782783	0.23205\\
0.283283283283283	0.23147\\
0.283783783783784	0.23098\\
0.284284284284284	0.23039\\
0.284784784784785	0.22983\\
0.285285285285285	0.22932\\
0.285785785785786	0.22863\\
0.286286286286286	0.22818\\
0.286786786786787	0.22754\\
0.287287287287287	0.22684\\
0.287787787787788	0.22628\\
0.288288288288288	0.22571\\
0.288788788788789	0.22523\\
0.289289289289289	0.22466\\
0.28978978978979	0.22416\\
0.29029029029029	0.22358\\
0.290790790790791	0.2231\\
0.291291291291291	0.22267\\
0.291791791791792	0.22197\\
0.292292292292292	0.22142\\
0.292792792792793	0.22092\\
0.293293293293293	0.2202\\
0.293793793793794	0.21955\\
0.294294294294294	0.21896\\
0.294794794794795	0.21846\\
0.295295295295295	0.21793\\
0.295795795795796	0.21733\\
0.296296296296296	0.21684\\
0.296796796796797	0.21624\\
0.297297297297297	0.21575\\
0.297797797797798	0.21532\\
0.298298298298298	0.21483\\
0.298798798798799	0.21436\\
0.299299299299299	0.21371\\
0.2997997997998	0.21312\\
0.3003003003003	0.21256\\
0.300800800800801	0.21196\\
0.301301301301301	0.21141\\
0.301801801801802	0.21083\\
0.302302302302302	0.21018\\
0.302802802802803	0.20963\\
0.303303303303303	0.2092\\
0.303803803803804	0.20864\\
0.304304304304304	0.20811\\
0.304804804804805	0.20751\\
0.305305305305305	0.20704\\
0.305805805805806	0.20651\\
0.306306306306306	0.20593\\
0.306806806806807	0.20537\\
0.307307307307307	0.20482\\
0.307807807807808	0.20434\\
0.308308308308308	0.2037\\
0.308808808808809	0.20321\\
0.309309309309309	0.20256\\
0.30980980980981	0.20203\\
0.31031031031031	0.20151\\
0.310810810810811	0.20092\\
0.311311311311311	0.2003\\
0.311811811811812	0.19979\\
0.312312312312312	0.1993\\
0.312812812812813	0.19884\\
0.313313313313313	0.19824\\
0.313813813813814	0.19768\\
0.314314314314314	0.19699\\
0.314814814814815	0.19633\\
0.315315315315315	0.19568\\
0.315815815815816	0.19507\\
0.316316316316316	0.19446\\
0.316816816816817	0.19378\\
0.317317317317317	0.19324\\
0.317817817817818	0.19263\\
0.318318318318318	0.19215\\
0.318818818818819	0.19157\\
0.319319319319319	0.19098\\
0.31981981981982	0.19046\\
0.32032032032032	0.18985\\
0.320820820820821	0.18943\\
0.321321321321321	0.18886\\
0.321821821821822	0.18833\\
0.322322322322322	0.18768\\
0.322822822822823	0.18722\\
0.323323323323323	0.18667\\
0.323823823823824	0.18613\\
0.324324324324324	0.18537\\
0.324824824824825	0.18477\\
0.325325325325325	0.18418\\
0.325825825825826	0.18366\\
0.326326326326326	0.18316\\
0.326826826826827	0.18259\\
0.327327327327327	0.18207\\
0.327827827827828	0.18142\\
0.328328328328328	0.18081\\
0.328828828828829	0.1802\\
0.329329329329329	0.17965\\
0.32982982982983	0.17914\\
0.33033033033033	0.17863\\
0.330830830830831	0.17811\\
0.331331331331331	0.17748\\
0.331831831831832	0.17689\\
0.332332332332332	0.17626\\
0.332832832832833	0.1757\\
0.333333333333333	0.17532\\
0.333833833833834	0.17485\\
0.334334334334334	0.17443\\
0.334834834834835	0.17378\\
0.335335335335335	0.17318\\
0.335835835835836	0.17265\\
0.336336336336336	0.17219\\
0.336836836836837	0.17148\\
0.337337337337337	0.17097\\
0.337837837837838	0.17025\\
0.338338338338338	0.16966\\
0.338838838838839	0.16909\\
0.339339339339339	0.16857\\
0.33983983983984	0.16795\\
0.34034034034034	0.16727\\
0.340840840840841	0.16681\\
0.341341341341341	0.1664\\
0.341841841841842	0.16587\\
0.342342342342342	0.16538\\
0.342842842842843	0.16489\\
0.343343343343343	0.16445\\
0.343843843843844	0.16378\\
0.344344344344344	0.16314\\
0.344844844844845	0.16263\\
0.345345345345345	0.16198\\
0.345845845845846	0.16147\\
0.346346346346346	0.16095\\
0.346846846846847	0.16037\\
0.347347347347347	0.15995\\
0.347847847847848	0.1594\\
0.348348348348348	0.15883\\
0.348848848848849	0.15826\\
0.349349349349349	0.15771\\
0.34984984984985	0.15712\\
0.35035035035035	0.15652\\
0.350850850850851	0.15592\\
0.351351351351351	0.15524\\
0.351851851851852	0.15471\\
0.352352352352352	0.15423\\
0.352852852852853	0.15359\\
0.353353353353353	0.15308\\
0.353853853853854	0.15239\\
0.354354354354354	0.15172\\
0.354854854854855	0.15111\\
0.355355355355355	0.1505\\
0.355855855855856	0.15012\\
0.356356356356356	0.14955\\
0.356856856856857	0.14917\\
0.357357357357357	0.14861\\
0.357857857857858	0.14806\\
0.358358358358358	0.14737\\
0.358858858858859	0.14692\\
0.359359359359359	0.14629\\
0.35985985985986	0.14578\\
0.36036036036036	0.14513\\
0.360860860860861	0.14462\\
0.361361361361361	0.14408\\
0.361861861861862	0.14362\\
0.362362362362362	0.14308\\
0.362862862862863	0.14242\\
0.363363363363363	0.14179\\
0.363863863863864	0.14125\\
0.364364364364364	0.14064\\
0.364864864864865	0.1401\\
0.365365365365365	0.13965\\
0.365865865865866	0.13923\\
0.366366366366366	0.1386\\
0.366866866866867	0.13807\\
0.367367367367367	0.13758\\
0.367867867867868	0.13711\\
0.368368368368368	0.1365\\
0.368868868868869	0.13592\\
0.369369369369369	0.13537\\
0.36986986986987	0.13478\\
0.37037037037037	0.13436\\
0.370870870870871	0.13368\\
0.371371371371371	0.13318\\
0.371871871871872	0.13252\\
0.372372372372372	0.13204\\
0.372872872872873	0.13157\\
0.373373373373373	0.13097\\
0.373873873873874	0.13043\\
0.374374374374374	0.12986\\
0.374874874874875	0.12938\\
0.375375375375375	0.12872\\
0.375875875875876	0.12825\\
0.376376376376376	0.12767\\
0.376876876876877	0.1271\\
0.377377377377377	0.12662\\
0.377877877877878	0.12614\\
0.378378378378378	0.12565\\
0.378878878878879	0.12513\\
0.379379379379379	0.12461\\
0.37987987987988	0.12401\\
0.38038038038038	0.12348\\
0.380880880880881	0.12303\\
0.381381381381381	0.1225\\
0.381881881881882	0.12184\\
0.382382382382382	0.12136\\
0.382882882882883	0.12084\\
0.383383383383383	0.12028\\
0.383883883883884	0.11968\\
0.384384384384384	0.11916\\
0.384884884884885	0.11863\\
0.385385385385385	0.1181\\
0.385885885885886	0.11755\\
0.386386386386386	0.11707\\
0.386886886886887	0.11659\\
0.387387387387387	0.11608\\
0.387887887887888	0.11554\\
0.388388388388388	0.11515\\
0.388888888888889	0.11455\\
0.389389389389389	0.11395\\
0.38988988988989	0.11349\\
0.39039039039039	0.11293\\
0.390890890890891	0.11246\\
0.391391391391391	0.11188\\
0.391891891891892	0.11126\\
0.392392392392392	0.11072\\
0.392892892892893	0.11023\\
0.393393393393393	0.10974\\
0.393893893893894	0.10912\\
0.394394394394394	0.1085\\
0.394894894894895	0.10795\\
0.395395395395395	0.10752\\
0.395895895895896	0.10697\\
0.396396396396396	0.10647\\
0.396896896896897	0.10586\\
0.397397397397397	0.10526\\
0.397897897897898	0.10464\\
0.398398398398398	0.1041\\
0.398898898898899	0.10357\\
0.399399399399399	0.10312\\
0.3998998998999	0.10267\\
0.4004004004004	0.10215\\
0.400900900900901	0.10167\\
0.401401401401401	0.10112\\
0.401901901901902	0.10061\\
0.402402402402402	0.10009\\
0.402902902902903	0.09946\\
0.403403403403403	0.09902\\
0.403903903903904	0.09845\\
0.404404404404404	0.09798\\
0.404904904904905	0.09747\\
0.405405405405405	0.09701\\
0.405905905905906	0.09639\\
0.406406406406406	0.09583\\
0.406906906906907	0.0953\\
0.407407407407407	0.09474\\
0.407907907907908	0.09422\\
0.408408408408408	0.09376\\
0.408908908908909	0.09319\\
0.409409409409409	0.0927\\
0.40990990990991	0.0922\\
0.41041041041041	0.09171\\
0.410910910910911	0.09114\\
0.411411411411411	0.09066\\
0.411911911911912	0.09017\\
0.412412412412412	0.08959\\
0.412912912912913	0.08905\\
0.413413413413413	0.08857\\
0.413913913913914	0.08805\\
0.414414414414414	0.08743\\
0.414914914914915	0.0869\\
0.415415415415415	0.08633\\
0.415915915915916	0.08571\\
0.416416416416416	0.08524\\
0.416916916916917	0.08464\\
0.417417417417417	0.08406\\
0.417917917917918	0.08356\\
0.418418418418418	0.08307\\
0.418918918918919	0.08247\\
0.419419419419419	0.08198\\
0.41991991991992	0.08144\\
0.42042042042042	0.08098\\
0.420920920920921	0.08052\\
0.421421421421421	0.08019\\
0.421921921921922	0.07978\\
0.422422422422422	0.07932\\
0.422922922922923	0.0788\\
0.423423423423423	0.07823\\
0.423923923923924	0.07779\\
0.424424424424424	0.07726\\
0.424924924924925	0.07678\\
0.425425425425425	0.07624\\
0.425925925925926	0.07563\\
0.426426426426426	0.07512\\
0.426926926926927	0.0746\\
0.427427427427427	0.07408\\
0.427927927927928	0.07349\\
0.428428428428428	0.07302\\
0.428928928928929	0.07255\\
0.429429429429429	0.07204\\
0.42992992992993	0.07152\\
0.43043043043043	0.07103\\
0.430930930930931	0.07052\\
0.431431431431431	0.07012\\
0.431931931931932	0.06973\\
0.432432432432432	0.06902\\
0.432932932932933	0.06848\\
0.433433433433433	0.06786\\
0.433933933933934	0.06735\\
0.434434434434434	0.06684\\
0.434934934934935	0.06644\\
0.435435435435435	0.06588\\
0.435935935935936	0.06541\\
0.436436436436436	0.06491\\
0.436936936936937	0.0644\\
0.437437437437437	0.06379\\
0.437937937937938	0.06332\\
0.438438438438438	0.06279\\
0.438938938938939	0.06222\\
0.439439439439439	0.06168\\
0.43993993993994	0.06118\\
0.44044044044044	0.06076\\
0.440940940940941	0.06032\\
0.441441441441441	0.05983\\
0.441941941941942	0.05925\\
0.442442442442442	0.05887\\
0.442942942942943	0.05846\\
0.443443443443443	0.05804\\
0.443943943943944	0.05761\\
0.444444444444444	0.0571\\
0.444944944944945	0.0566\\
0.445445445445445	0.05608\\
0.445945945945946	0.05555\\
0.446446446446446	0.05502\\
0.446946946946947	0.05445\\
0.447447447447447	0.05394\\
0.447947947947948	0.0535\\
0.448448448448448	0.05289\\
0.448948948948949	0.05242\\
0.449449449449449	0.0519\\
0.44994994994995	0.05143\\
0.45045045045045	0.05095\\
0.450950950950951	0.05034\\
0.451451451451451	0.04977\\
0.451951951951952	0.04913\\
0.452452452452452	0.04871\\
0.452952952952953	0.04813\\
0.453453453453453	0.04743\\
0.453953953953954	0.04671\\
0.454454454454454	0.04624\\
0.454954954954955	0.04568\\
0.455455455455455	0.04525\\
0.455955955955956	0.04462\\
0.456456456456456	0.04412\\
0.456956956956957	0.04358\\
0.457457457457457	0.04316\\
0.457957957957958	0.04261\\
0.458458458458458	0.04204\\
0.458958958958959	0.04148\\
0.459459459459459	0.04105\\
0.45995995995996	0.0406\\
0.46046046046046	0.04018\\
0.460960960960961	0.03975\\
0.461461461461461	0.03923\\
0.461961961961962	0.03867\\
0.462462462462462	0.03812\\
0.462962962962963	0.03765\\
0.463463463463463	0.03716\\
0.463963963963964	0.03654\\
0.464464464464464	0.0361\\
0.464964964964965	0.03555\\
0.465465465465465	0.03502\\
0.465965965965966	0.03461\\
0.466466466466466	0.03413\\
0.466966966966967	0.03358\\
0.467467467467467	0.03307\\
0.467967967967968	0.03259\\
0.468468468468468	0.03191\\
0.468968968968969	0.03138\\
0.469469469469469	0.03096\\
0.46996996996997	0.03039\\
0.47047047047047	0.02993\\
0.470970970970971	0.02923\\
0.471471471471471	0.02883\\
0.471971971971972	0.02833\\
0.472472472472472	0.0278\\
0.472972972972973	0.02732\\
0.473473473473473	0.02685\\
0.473973973973974	0.02643\\
0.474474474474474	0.02588\\
0.474974974974975	0.02542\\
0.475475475475475	0.02487\\
0.475975975975976	0.02433\\
0.476476476476476	0.02389\\
0.476976976976977	0.02331\\
0.477477477477477	0.02279\\
0.477977977977978	0.02235\\
0.478478478478478	0.02191\\
0.478978978978979	0.02142\\
0.479479479479479	0.02092\\
0.47997997997998	0.02036\\
0.48048048048048	0.01991\\
0.480980980980981	0.01943\\
0.481481481481481	0.01904\\
0.481981981981982	0.01847\\
0.482482482482482	0.01797\\
0.482982982982983	0.01743\\
0.483483483483483	0.01683\\
0.483983983983984	0.01643\\
0.484484484484485	0.0159\\
0.484984984984985	0.01539\\
0.485485485485485	0.01492\\
0.485985985985986	0.01442\\
0.486486486486487	0.01408\\
0.486986986986987	0.01357\\
0.487487487487487	0.01308\\
0.487987987987988	0.0126\\
0.488488488488488	0.01203\\
0.488988988988989	0.01157\\
0.48948948948949	0.01102\\
0.48998998998999	0.0104\\
0.49049049049049	0.00976\\
0.490990990990991	0.00921\\
0.491491491491492	0.00874\\
0.491991991991992	0.00824\\
0.492492492492492	0.0077\\
0.492992992992993	0.00715\\
0.493493493493493	0.0065\\
0.493993993993994	0.00601\\
0.494494494494495	0.00555\\
0.494994994994995	0.00526\\
0.495495495495495	0.00482\\
0.495995995995996	0.00424\\
0.496496496496497	0.00375\\
0.496996996996997	0.00312\\
0.497497497497497	0.00259\\
0.497997997997998	0.002\\
0.498498498498498	0.00145\\
0.498998998998999	0.00089\\
0.4994994994995	0.00045\\
0.5	0\\
};
\end{axis}
\end{tikzpicture}%

%% file: main.bbl
\begin{thebibliography}{10}
\providecommand{\url}[1]{#1}
\csname url@samestyle\endcsname
\providecommand{\newblock}{\relax}
\providecommand{\bibinfo}[2]{#2}
\providecommand{\BIBentrySTDinterwordspacing}{\spaceskip=0pt\relax}
\providecommand{\BIBentryALTinterwordstretchfactor}{4}
\providecommand{\BIBentryALTinterwordspacing}{\spaceskip=\fontdimen2\font plus
\BIBentryALTinterwordstretchfactor\fontdimen3\font minus
  \fontdimen4\font\relax}
\providecommand{\BIBforeignlanguage}[2]{{%
\expandafter\ifx\csname l@#1\endcsname\relax
\typeout{** WARNING: IEEEtran.bst: No hyphenation pattern has been}%
\typeout{** loaded for the language `#1'. Using the pattern for}%
\typeout{** the default language instead.}%
\else
\language=\csname l@#1\endcsname
\fi
#2}}
\providecommand{\BIBdecl}{\relax}
\BIBdecl

\bibitem{linnartz2003new}
J.-P. Linnartz and P.~Tuyls, ``{New Shielding Functions to Enhance Privacy and
  Prevent Misuse of Biometric Templates},'' in \emph{Audio- and Video-based
  Biometric Person Auth.}\hskip 1em plus 0.5em minus 0.4em\relax Springer,
  2003, pp. 393--402.

\bibitem{dodis2004fuzzy}
Y.~Dodis, L.~Reyzin, and A.~Smith, ``{Fuzzy Extractors: How to Generate Strong
  Keys from Biometrics and other Noisy Data},'' in \emph{Advances in
  Cryptology-Eurocrypt}.\hskip 1em plus 0.5em minus 0.4em\relax Springer, 2004,
  pp. 523--540.

\bibitem{hiller2013breaking}
M.~Hiller, M.~Weiner, L.~Rodrigues~Lima, M.~Birkner, and G.~Sigl, ``{Breaking
  Through Fixed PUF Block Limitations with Differential Sequence Coding and
  Convolutional Codes},'' in \emph{TrustED}.\hskip 1em plus 0.5em minus
  0.4em\relax ACM, 2013, pp. 43--54.

\bibitem{muelich2016new}
S.~M{\"u}elich and M.~Bossert, ``{A New Error Correction Scheme for Physical
  Unclonable Functions},'' in \emph{ITG SCC}, 2017.

\bibitem{bosch2008efficient}
C.~B{\"o}sch, J.~Guajardo, A.-R. Sadeghi, J.~Shokrollahi, and P.~Tuyls,
  ``Efficient helper data key extractor on fpgas,'' \emph{Cryptographic
  Hardware and Embedded Systems--CHES 2008}, pp. 181--197, 2008.

\bibitem{maes2012pufky}
R.~Maes, A.~Van~Herrewege, and I.~Verbauwhede, ``Pufky: A fully functional
  puf-based cryptographic key generator,'' \emph{Cryptographic Hardware and
  Embedded Systems--CHES 2012}, pp. 302--319, 2012.

\bibitem{muelich2014error}
S.~M{\"u}elich, S.~Puchinger, M.~Bossert, M.~Hiller, and G.~Sigl, ``Error
  correction for physical unclonable functions using generalized concatenated
  codes,'' \emph{Int. Workshop on Algebraic and Combinatorial Coding Theory},
  2014.

\bibitem{puchinger2015error}
S.~Puchinger, S.~M{\"u}elich, M.~Bossert, M.~Hiller, and G.~Sigl, ``On error
  correction for physical unclonable functions,'' in \emph{SCC 2015; 10th
  International ITG Conference on Systems, Communications and Coding;
  Proceedings of}.\hskip 1em plus 0.5em minus 0.4em\relax VDE, 2015, pp. 1--6.

\bibitem{hiller2016}
M.~Hiller, A.~G. {\"O}nalan, G.~Sigl, and M.~Bossert, ``{Online Reliability
  Testing for PUF Key Derivation},'' in \emph{Proc. of the 6th Int. Workshop on
  Trustworthy Embedded Devices}.\hskip 1em plus 0.5em minus 0.4em\relax ACM,
  2016, pp. 15--22.

\bibitem{maes2009low}
R.~Maes, P.~Tuyls, and I.~Verbauwhede, ``Low-overhead implementation of a soft
  decision helper data algorithm for sram pufs.'' in \emph{CHES}, vol.~9.\hskip
  1em plus 0.5em minus 0.4em\relax Springer, 2009, pp. 332--347.

\bibitem{guajardo2007fpga}
J.~Guajardo, S.~S. Kumar, G.-J. Schrijen, and P.~Tuyls, ``{FPGA Intrinsic PUFs
  and their Use for IP Protection},'' in \emph{CHES}.\hskip 1em plus 0.5em
  minus 0.4em\relax Springer, 2007, pp. 63--80.

\bibitem{maes2009soft}
R.~Maes, P.~Tuyls, and I.~Verbauwhede, ``{A Soft Decision Helper Data Algorithm
  for SRAM PUFs},'' in \emph{IEEE ISIT}.\hskip 1em plus 0.5em minus 0.4em\relax
  IEEE, 2009, pp. 2101--2105.

\bibitem{elias1955coding}
P.~Elias, ``{Coding for Noisy Channels},'' in \emph{Proceedings of the
  Institute of Radio Engineers}, no.~3, 1955, pp. 37--46.

\bibitem{hiller2014seesaw}
M.~Hiller, L.~R. Lima, and G.~Sigl, ``{Seesaw: An Area-Optimized FPGA Viterbi
  Decoder for PUFs},'' in \emph{17th Euromicro Conference on Digital System
  Design}.\hskip 1em plus 0.5em minus 0.4em\relax IEEE, 2014, pp. 387--393.

\bibitem{costello2004error}
D.~Costello and S.~Lin, ``{Error Control Coding},'' \emph{New Jersey}, 2004.

\bibitem{johannesson1999fundamentals}
R.~Johannesson and K.~Zigangirov, ``{Fundamentals of Convolutional Codes},''
  1999.

\bibitem{viterbi1967error}
A.~Viterbi, ``{Error Bounds for Convolutional Codes and an Asymptotically
  Optimum Decoding Algorithm},'' \emph{IEEE Transactions on Information
  Theory}, vol.~13, no.~2, pp. 260--269, 1967.

\bibitem{schmidt2004finding}
G.~Schmidt, V.~Sidorenko, V.~V. Zyablov, and M.~Bossert, ``{Finding a List of
  Best Paths in a Trellis},'' in \emph{IEEE ISIT}, 2004.

\bibitem{wozencraft1957sequential}
J.~M. Wozencraft, ``{Sequential Decoding for Reliable Communication},'' 1957.

\bibitem{zigangirov1966some}
K.~Zigangirov, ``{Some Sequential Decoding Procedures},'' \emph{Problemy
  Peredachi Informatsii}, vol.~2, no.~4, pp. 13--25, 1966.

\bibitem{jelinek1969fast}
F.~Jelinek, ``{A Fast Sequential Decoding Algorithm Using a Stack},'' \emph{IBM
  Journal of Research and Development}, vol.~13, no.~6, pp. 675--685, 1969.

\bibitem{fano1963heuristic}
R.~Fano, ``{A Heuristic Discussion of Probabilistic Decoding},'' \emph{IEEE
  Transactions on Information Theory}, vol.~9, no.~2, pp. 64--74, 1963.

\bibitem{van2012soft}
V.~Van~der Leest, B.~Preneel, and E.~Van~der Sluis, ``Soft decision error
  correction for compact memory-based pufs using a single enrollment.'' in
  \emph{CHES}.\hskip 1em plus 0.5em minus 0.4em\relax Springer, 2012, pp.
  268--282.

\end{thebibliography}
